\documentclass{sig-alt-release2}
\usepackage{pgfplots}
\usepackage{titlesec}
\usepackage{booktabs}
\titlespacing*{\section}{0pt}{5pt}{4pt}

\begin{document}

\title{Recency Ranking by Diversification of Result Set}

\conferenceinfo{CIKM'11,} {October 24--28, 2011, Glasgow, Scotland, UK.}
\CopyrightYear{2011}
\crdata{978-1-4503-0717-8/11/10}
\clubpenalty=10000
\widowpenalty = 10000

%\subtitle{[Extended Abstract]
%\titlenote{A full version of this paper is available as
%\textit{Author's Guide to Preparing ACM SIG Proceedings Using
%\LaTeX$2_\epsilon$\ and BibTeX} at
% You need the command \numberofauthors to handle the "boxing"
% and alignment of the authors under the title, and to add
% a section for authors number 4 through n.
%
% Up to the first three authors are aligned under the title;
% use the \alignauthor commands below to handle those names
% and affiliations. Add names, affiliations, addresses for
% additional authors as the argument to \additionalauthors;
% these will be set for you without further effort on you-l=-  % part as the last section in the body of your article BEFORE
% References or any Appendices.
\numberofauthors{1} %  in this sample file, there are a *total*

%\author{Submitted for Confidential Blind Review}

%\institute{Anonymous\\}

\author{\alignauthor Andrey Styskin, Fedor Romanenko, Fedor Vorobyev, Pavel Serdyukov\\
       \affaddr{Search Quality Group, Yandex}\\
       \affaddr{119021, Leo Tolstoy 16}\\
       \affaddr{Moscow, Russia}\\
       \email{\{styskin, fedor, melton, pavser\}@yandex-team.ru}
}

\maketitle 
\begin{abstract}
In this paper, we propose a web search retrieval approach which automatically detects recency sensitive queries and increases the freshness of the ordinary document ranking by a degree proportional to the probability of the need in recent content. We propose to solve the recency ranking problem by using result diversification principles and deal with the query's non-topical ambiguity appearing when the need in recent content can be detected only with uncertainty. Our offline and online experiments with millions of queries from real search engine users demonstrate the significant increase in satisfaction of users presented with a search result generated by our approach.

\end{abstract}

% A category with only the three required fields
\normalsize
\vspace{1mm}
\noindent
{\bf Categories and Subject Descriptors: } \\H.3 [Information Storage and Retrieval]:
H.3.3 Information Search and Retrieval.

\vspace{1mm}
\noindent
{\bf General Terms:\\
\normalfont
Algorithms, Measurement, Performance, Experimentation.}

\vspace{1mm}
\noindent
{\bf Keywords:\\
\normalfont
Recency ranking, diversity, web search.}

\section{Introduction}

Modern web search engines face the need to consider different non-topical facets of relevance when ranking web documents in response to user queries. While the need in a certain document feature, besides the topical relevance, might be expressed only implicitly in the query, it is still important to recognize its presence in order to adequately satisfy the underlying information need. However, the quality of such recognition cannot be perfect in all cases. Consequently, it brings a certain level of non-topical ambiguity to the queries, which must be taken into account when generating a search result. \par
One of the most popular non-topical facets of relevance is document freshness. In this paper, we demonstrate how to deal with query ambiguity surrounding the need for recent information. Such ambiguity typically appears during short periods of time, when users become increasingly interested in one newsworthy aspect, typically an event, related to a well-known entity: a person (e.g.\ [Michael Jackson]) or a location (e.g.\ [Japan]). For example, consider the query [michael jackson].  There has been constant and continuing interest in the biography and discography of Michael Jackson, which can be satisfied even by non-fresh documents, but users issuing the query [michael jackson] on (June 25, 2009) were often looking for the freshest news related to the Michael Jackson's death.  But both intents (news and discography) continued even during this period. \par
Recently, a number of techniques for search result diversification has been proposed in order to compensate for topical ambiguity of queries and increase the chance to satisfy the user. We propose to follow the same principles to deal with recency sensitive queries and their non-topical ambiguity. Our approach aims to maximize the probability that the average user finds some useful information among the search results on recency sensitive queries by blending necessary amount of recent results into the result set. To the best of our knowledge, our paper describes the first attempt to tackle the problem of non-topical query ambiguity with a result set diversification technique. \par
The main contributions of this paper include: 1) the approach to recency ranking by means of search result set diversification, 2) a thorough offline and online evaluation of the proposed approach in terms of a search result quality metric and the overall user satisfaction.  \par
The remainder of the paper is organized as follows. We first review the related work in Section 2. Section 3 describes our machine learning based approach to obtaining a smooth probability of the need in recent content for a query. Section 4 explains how we utilize that probability to diversify ordinary document rankings with fresh documents. Section 5 presents the results of evaluation and Section 6 concludes the paper and outlines the research questions left for future work. \par
%The naive approach was criticized by Amit Singhal -- the head of Google's core ranking team -- in his interview for New York Times "Google Keeps Tweaking Its Search Engine By Saul Hansell". He explained that simply changing formulas to display more new pages results in lower quality searches much of the time. \par

\section{RELATED WORK}
\label{related}
% Recent ranking works

There are only a few papers focused exclusively on recency ranking. The pioneering work in this area proposed to learn a ranking function which is trained using a subset of features that help to infer the recency of page content \cite{TowardsRR}. The follow-up work extended that subset to include features extracted from the micro-blogging data stream \cite{TowardsRR2}. Our approach differs in a number of aspects. For one, we try to deal with temporal ambiguity of queries and balance the number of fresh and ordinary relevant documents in the result set, based on the smooth probability of the need in recent content. As a result, while the aforementioned works focus entirely on improving the ranking for one specific case (breaking news queries, 1-2\% of search engine's traffic), we aim to affect the ordinary ranking by explicitly increasing its freshness for any query with non-zero probability of the need in recent content. The inference of query's recency sensitivity plays an important role in recency ranking. The aforementioned works detected only highly recency sensitive queries using a linear combination of a few features. In a similar way, Arguello et.\ al.\ \cite{NewsRR} used a number of features to find verticals relevant for a query, including the News vertical. While those methods focused on binary classification of a query, our work is rather based on regression to obtain and utilize the precise estimate of the probability of the need in recent content.

% Diversity works
Our work is also largely based on the principles of search result diversification.  In \cite{SIGIR08}, a framework for evaluation that systematically rewards diversity was proposed. In \cite{WSDM09}, a systematic approach to diversifying results that aims to minimize the risk of dissatisfaction of the average user was presented. A number of follow-up papers were published recently which we do not cover here due to space constraints. Our work complements the research in this area by demonstrating that diversification principles and algorithms are also helpful to increase the chance to satisfy the user in the presence of non-topical query ambiguity. 

% DIR works
%The problem of integration news results is related to distributed information retrieval that refers to the situation where a searcher's query is satisfied by retrieving content from various sub-collections \cite{DIR}.  There are three aspects in DIR: resource description, resource selection, and results merging. The main approach to merge result is to normalize relevance scores in different collections as discussed in \cite{Normalization}. In our work we operate with two collections: recent documents collection and big web collection. There is no straightforward way to normalize relevance of these collections because of different nature of needs in documents from these collections. So, we propose way of merging results based on diversification model. \par

\section{RECENCY SENSITIVE QUERY \\ CLASSIFICATION}
\label{queryclassifier}

In order to quantify the ambiguity of a recency sensitive query, we learned a regression model which predicted the level of interest in recent documents for a particular query and a particular time slot. We used around 30 different features (including their minor modifications) previously described in the papers dealing with similar problems (see Section \ref{related}). We do not provide a thorough analysis of feature importance due to space constraints. The most valuable features were the probabilities of queries to be generated by language models of recent content from different sources, including the query, social and news data streams, as well as the probability of a click on a news item. 

To train the regression model we asked annotators to provide labels of recency sensitivity for a set of queries. In order to preselect a list of candidate queries for assessment, we defined a small threshold on each feature used to learn our regression function and filtered out all queries that did not have at least one feature value exceeding the corresponding threshold. 
%We made test judgment and figured out its precision of 10\%. It leaved 30\% of search engine query traffic. 
As a result, we collected judgments for a set of 4000 unique queries issued to Yandex (www.yandex.ru) web search engine (the major russian search engine) over the period of three weeks. On each day during this period, judges were presented with the queries submitted by search engine users on that particular day and were asked to determine whether these queries express an interest in upcoming or ongoing events for which web search users would prefer recent content. Labeling queries basically represented the manual assignment of the probabilities that a particular query is recency sensitive. So, if a query was strictly about a recent event it received the probability of 0.95 (e.g.\ [flood in thailand] on the day of the event). If a query's primary interest was related to a recent event, but many users would also like to see just topically relevant results, it was labeled with 0.75 (e.g.\ the query [oscar] on the day of the ceremony). If the query's primary interest was not likely to be focused on a particular event, but there was some chance that users issuing such a query would look for some fresh content, assessors assigned the probability of 0.25 (e.g.\ it always makes sense to present users with some recent content in response to the query [britney spears]). Otherwise, a query was assigned zero probability to be recency sensitive. Each query was labeled by 3 assessors. Average Cohen's kappa coefficient between all pairs of assessors was 0.76, which is considered a substantial agreement. 

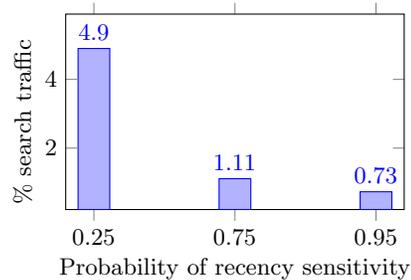
\begin{figure}
\centering
\begin{tikzpicture}
\begin{axis}[
height=2.6cm,
width=4.5cm,
scale only axis,
xlabel=Probability of recency sensitivity,
xticklabels={$0$,$1$,$0.25$,$0.75$,$0.95$},
ylabel=\% search traffic,
y label style={at={(axis description cs:0.2,.45)},anchor=south},
ymax=5.9,
ybar=13pt,
bar width=12pt,
nodes near coords
]
\addplot
coordinates { 
(1, 4.9) 
(2, 1.11) 
(3, 0.73)
%(0.25, 4.9) 
%(0.75, 1.11) 
%(0.95, 0.73)
};
\end{axis}
\end{tikzpicture}
\caption{\label{fig:coverage} Recency sensitive queries traffic coverage}
\end{figure}

We learned the regression model to assign smooth probability of the need in fresh content to any query using Gradient Boosted Regression Trees (GDBT) \cite{GBoost}. Recency sensitive queries traffic coverage by these types based on 4000 human made judgments is illustrated on Figure \ref{fig:coverage}.

%We report square roots of mean square errors for  in \ref{tab:marks}.
%\begin{table}[h]
%\caption{\label{tab:marks} Time sensitive class and probability equivalence}
%\begin{center}
%\begin{tabular}{|c|c|}
%\hline Recent Result Need Probability & $\sigma$ \\
%\hline 0 & 0.12 \\
%\hline 0.25 & 0.11 \\
%\hline 0.75 & 0.47\\
%\hline 0.95 & 0.57\\
%\end{tabular}
%\end{center}
%\end{table}

\section{Diversification of the search \\ result with fresh documents}

\subsection{Diversification method}

To produce a search result for recency sensitive queries we follow the search result set diversification principles. Namely, we aim to maximize the utility of the diversified search result expressed in terms of the Expected Reciprocal Rank measure, which we extended to include an abandonment probability and to handle multiple query intents. Both extensions are proposed by Chapelle et.\ al.\ \cite{ERR} and we combine both of them in this work. We call this metric Intent Aware Expected Reciprocal Rank with Abandonment (ERR-IAA) and regard as the objective we aim to maximize:

\begin{equation}
\label{diversify}
\text{ERR-IAA} = \sum^r_{i=1}{pBreak^r * \sum_t{P(t|q) * \prod^{r-1}_{i=1}{(1-R^t_i)} * R^t_r}},
\end{equation}

%\begin{equation}
%\prod^{r-1}_{i=1}{(1-R^t_i)} * R^t_r
%\end{equation}

%where $g^t_i$ denotes the grade of the document in position $i$ for topic $t$. The associated probability of relevance $R^t_i = \rho(g^t_i)$. 

where $P(t|q)$ is taken from the distribution over two classes of information needs $t$ (the need in fresh topically relevant documents and the need in any topically relevant documents) for the given query $q$. Each document is assigned the probability $R^t_r$ to satisfy the information need of type $t$ at position $r$. We take into account the probability that the documents at the previous $r-1$ positions have not satisfied that need. We also assume that any user always may stop (abandon the search result) at rank $r$ with the abandonment probability $pBreak^r$ due to accumulated frustration ($pBreak$ is empirically set to 0.85 in our experiments). \par
We assume that the optimal search result page is the one which maximizes the ERR-IAA measure. In order to maximize it, we follow the greedy approach described by Agrawal et.\ al.\ \cite{WSDM09} and select the document, whose selection leads to the maximum increase of the objective at each step of the selection process. 
\\

\subsection{Aggregation of ordinary and fresh results}

\begin{figure}
\centering
\begin{tikzpicture}
\begin{axis}[
height=3.2cm,
width=4cm,
scale only axis,
ymin=0,
ymax=130,
xlabel=Days,
ylabel=\% times,
y label style={at={(axis description cs:0.15,.45)},anchor=south},
ybar=13pt,
bar width=12pt,
nodes near coords,
point meta=explicit symbolic
]
\addplot
coordinates { 
(1, 73) [73\%]
(2, 20) [20\%]
(3, 4) [4\%]
};
\end{axis}
\end{tikzpicture}
\caption{\label{fig:3days} Cumulative share of query instances submitted since the day of the first query}
\end{figure}
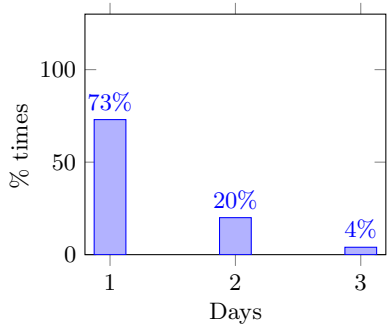

We also assume that any web document is fresh only for 3 days since the time of its creation or the last update. We use a proprietary algorithm to extract the correct and the most relevant timestamp from document content. Our choice of the number of days is motivated by the following observations. First, according to the studies by Dong et.\ al.\ \cite{TowardsRR}, assessors, who judged the freshness of a set of documents, found out that the 1--4 days old documents are the ones most likely containing fresh content. Second, the peak of interest in new events lasts for three days in average according to our analysis of 100,000 spiky and long-tail (so, previously unseen) queries submitted by users of Yandex search engine in January, 2011. Figure \ref{fig:3days} shows average share of total query volume for each such a query per for each day since the 1st day they become first known to the search engine until the 3rd day when their popularity falls off almost completely. 

\par

Guided by our definition of a fresh document, we produce the ranking of fresh topically relevant documents by simply removing the outdated documents from the initial ranking. As a result, we have two document rankings which we use to maximize the ERR-IAA measure, described by Equation \ref{diversify}: the one containing any topically relevant documents and the one containing only fresh topically relevant documents.

However, in order to proceed with maximization of our objective, we still needed to determine the probabilities of relevance $R^t_i$. In order to be independent of specific retrieval scores, which may significantly vary over queries and bear a relative, rather than absolute meaning, we turn document ranks into their probabilities of relevance using the internal search engine's statistics about the probability to encounter a relevant document at the specific position. So, since we fixed the probabilities of document relevance for each of two aggregated rankings, the final ordering of documents depended only on the probabilistic output of our classifier of query's recency sensitivity.

%However, as we considered a significant number of highly ranked documents to be outdated and removed them from the ranking of fresh topically relevant documents, we assumed that the overall relevance of fresh results might be lower than of general results. In order to validate our intuition and determine the rules to turn the document ranks into their probabilities of relevance, we asked a group of assessors to judge relevance of top-10 retrieved documents for 40 randomly sampled recency-sensitive queries. Based on these judgments we had fixed probabilities of relevance for the documents at ranks from 1 to 3 to 0.41, for the documents at ranks from 4 to 7 to 0.14 and for the documents at ranks from 8 to 10 to 0.07. For the rankings of fresh topically relevant documents we made only one change - the documents at ranks from 1 to 3 were in average less relevant and hence we assigned them the probability of only 0.3. So, since we fixed the probabilities of document relevance for each of two aggregated rankings, the final ordering of documents depended only on the probabilistic output of our classifier of query's recency sensitivity.

\section{Experiments}
\subsection{Offline results}

The research questions we aim to answer in this subsection are how our search quality objective (ERR-IAA) changes while we aggregate two result sets: ordinary and fresh, and how the quality of our classifier of query's recency sensitivity affects the quality of such aggregation. 

\begin{figure}
\centering
\begin{tikzpicture}
\begin{axis}[
height=7.5cm,
width=8.5cm,
legend style={at={(0.5,0.35)},anchor=north},
xlabel=Recent documents need approximation,
ylabel=ERR-IAA
]
\addplot[color=red,mark=x] coordinates {
( 0.0 , 0.578789572613 )
( 0.05 , 0.613937719176 )
( 0.1 , 0.641756214188 )
( 0.15 , 0.649409881814 )
( 0.2 , 0.650420545234 )
( 0.25 , 0.650420545234 )
( 0.3 , 0.650024151997 )
( 0.35 , 0.644773145713 )
( 0.4 , 0.643896285896 )
( 0.45 , 0.62200483725 )
( 0.5 , 0.613274826449 )
( 0.55 , 0.599302579574 )
( 0.6 , 0.564427579574 )
( 0.65 , 0.551100299418 )
( 0.7 , 0.518587799418 )
( 0.75 , 0.482920088095 )
( 0.8 , 0.477448603755 )
( 0.85 , 0.455399558903 )
( 0.9 , 0.398672910776 )
( 0.95 , 0.307251525959 )
};

\addplot[color=blue,mark=x] coordinates {
( 0.0 , 0.192929857538 )
( 0.05 , 0.306909997554 )
( 0.1 , 0.412243130986 )
( 0.15 , 0.449336488136 )
( 0.2 , 0.456195844036 )
( 0.25 , 0.456195844036 )
( 0.3 , 0.463478543464 )
( 0.35 , 0.492627872952 )
( 0.4 , 0.495271967891 )
( 0.45 , 0.534147111829 )
( 0.5 , 0.543626847537 )
( 0.55 , 0.554141931912 )
( 0.6 , 0.572516931912 )
( 0.65 , 0.577102142693 )
( 0.7 , 0.584114642693 )
( 0.75 , 0.584976828496 )
( 0.8 , 0.584710522281 )
( 0.85 , 0.581838342142 )
( 0.9 , 0.570029486934 )
( 0.95 , 0.546825086875 )
};

\addplot[color=green,mark=x] coordinates {
( 0.0 , 0.0385859715075 )
( 0.05 , 0.184098908905 )
( 0.1 , 0.320437897705 )
( 0.15 , 0.369307130665 )
( 0.2 , 0.378505963557 )
( 0.25 , 0.378505963557 )
( 0.3 , 0.388860300051 )
( 0.35 , 0.431769763847 )
( 0.4 , 0.435822240689 )
( 0.45 , 0.499004021661 )
( 0.5 , 0.515767655972 )
( 0.55 , 0.536077672847 )
( 0.6 , 0.575752672847 )
( 0.65 , 0.587502880004 )
( 0.7 , 0.610325380004 )
( 0.75 , 0.625799524657 )
( 0.8 , 0.627615289691 )
( 0.85 , 0.632413855437 )
( 0.9 , 0.638572117397 )
( 0.95 , 0.642654511241 )
};
\legend{ERR-IAA for 0.25, ERR-IAA for 0.75, ERR-IAA for 0.95}
\end{axis}
\end{tikzpicture}
\caption{\label{fig:single} ERR-IAA for queries with different ``true" probabilities of need in fresh content}
\end{figure}
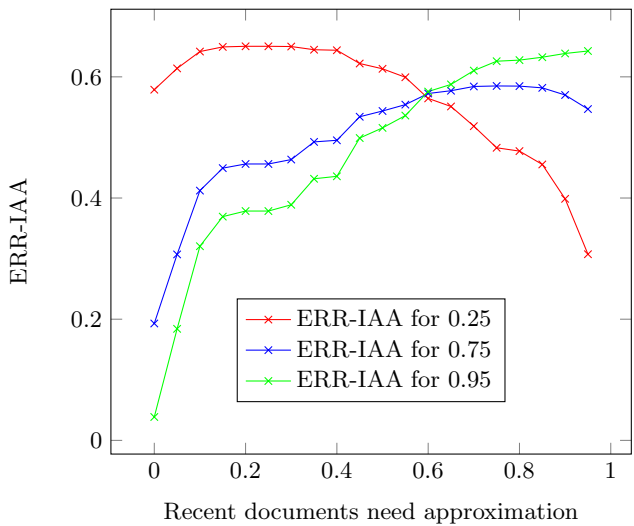

Figure \ref{fig:single} demonstrates how ERR-IAA changes as the estimation of the probability of the need in recent documents deviates from its true value for the three different true values assigned by our assessors: 0.25, 0.75 and 0.95. As we see, while minor errors in the probability estimation do not significantly affect the quality of the aggregated ranking, it is evidently important to keep the errors low as the ranking quality drops quite rapidly with their increase. \par

%Let's now assume that we have an ideal recency need classifier and a hypothesized situation, when the initial ranking does not include any fresh documents. The objective values that depend on the probability of the need in recent content are shown on Figure \ref{fig:derr}. 

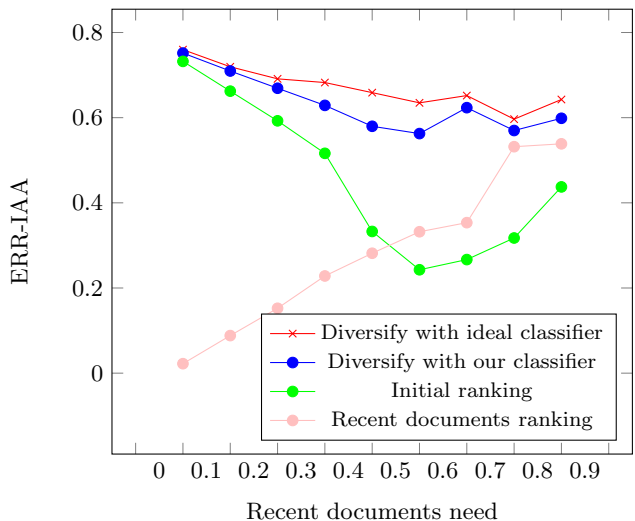
\begin{figure}
\centering
\begin{tikzpicture}
\begin{axis}[
height=7.5cm,
width=8.5cm,
legend style={font=\small},
legend pos=south east,
xtick={0,0.1,0.2,0.3,0.4,0.5,0.6,0.7,0.8,0.9,1},
x tick label as interval,
xlabel=Recent documents need,
ymin=-0.19,
xmin=-0.05,
xmax=1.05,
ylabel=ERR-IAA
]
\addplot[color=red,mark=x] coordinates {
( 0.1 , 0.75972529683 )
( 0.2 , 0.719495339085 )
( 0.3 , 0.691181390539 )
( 0.4 , 0.682493454075 )
( 0.5 , 0.658795806632 )
( 0.6 , 0.634831196401 )
( 0.7, 0.651903899691 )
( 0.8 , 0.596512365045 )
( 0.9 , 0.642654511241 )
};
\addplot[color=blue,mark=*] coordinates {
( 0.1 , 0.751697817574 )
( 0.2 , 0.709576901374 )
( 0.3 , 0.669097880089 )
( 0.4 , 0.628854860734 )
( 0.5 , 0.579759960163 )
( 0.6 , 0.562604622559 )
( 0.7 , 0.623478476331 )
( 0.8 , 0.569820840394 )
( 0.9 , 0.598551688894 )
};
\addplot[color=green,mark=*] coordinates {
( 0.1 , 0.732166889746 )
( 0.2 , 0.662288732915 )
( 0.3 , 0.592579738619 )
( 0.4 , 0.516240654038 )
( 0.5 , 0.332763243568 )
( 0.6 , 0.24279684911 )
( 0.7 , 0.266726163454 )
( 0.8 , 0.317412840656 )
( 0.9 , 0.437404399687 )
};

\addplot[color=pink,mark=*] coordinates {
( 0.1 , 0.0222420895989 )
( 0.2 , 0.0882203553825 )
( 0.3 , 0.152356403219 )
( 0.4 , 0.228215719333 )
( 0.5 , 0.281607896967 )
( 0.6 , 0.332097022623 )
( 0.7 , 0.353431423746 )
( 0.8 , 0.531830142398 )
( 0.9 , 0.538562169517)
};

\legend{Diversify with ideal classifier, Diversify with our classifier, Initial ranking, Recent documents ranking}
\end{axis}
\end{tikzpicture}
\caption{\label{fig:offline} ERR-IAA for queries with different level of ambiguity, diversified and non-diversified rankings}
\end{figure}

Further, we analyze the quality of the aggregation for the recency sensitivity classifier that we use in this work (Figure \ref{fig:offline}). We split our queries judged by their recency sensitivity (see Section \ref{queryclassifier}) into two parts (training and test, 2000 queries each) and conduct the evaluation via two-fold cross-validation. We train our classifier on the training set of queries and evaluate how its accuracy affects the quality of the aggregation on the test set. Note, that in a real setting that we simulate in this experiment, the initial ranking naturally contains some number of fresh documents. As a result, ERR-IAA measured on the ordinary ranking starts to grow as the recency need probability approaches 1.0, since the queries with such high probability of the need in recent content are typically the queries that are unambiguous: highly descriptive and possessing enough discriminating power to retrieve very relevant content. For example, for the query [europe alert icelandic ash cloud], both the ordinary result set and the fresh result set are quite similar on the day of the infamous volcano eruption. The major gain from applying the diversification comes for the queries with probabilities of the need in recent content from 0.3 to 0.8. This is to be expected, as our approach to recency ranking focuses on the cases of temporal query ambiguity, in contrast to the previous approaches, which aim to maximize the quality of ranking for the queries with no temporal ambiguity (see \cite{TowardsRR,TowardsRR2} for more details).

%Average ideal metric on the training set is 0.73, average metric based on our classifier -- 0.71, metric for actual result set -- 0.66. The objective values for our classifier are shown on Figure \ref{fig:offline}.  \par

%\begin{comment}
%\begin{figure}
%\centering
%\begin{tikzpicture}
%\begin{axis}[
%xtick=data,
%xlabel=Recent documents need,
%ylabel=Metric with ideal classifier]
%\addplot[color=red,mark=x] coordinates {
%( 0, 0.77171943)
%( 0.25 , 0.65042)
%( 0.75 , 0.58497)
%( 0.95 , 0.6426)
%};
%\addplot[color=blue,mark=*] coordinates {
%( 0.0 , 0.771)
%( 0.25 , 0.57878)
%( 0.75 , 0.192929857538)
%( 0.95, 0.03858)
%};
%\legend{With recent documents, without recent document}
%\end{axis}
%\end{tikzpicture}
%\caption{\label{fig:derr} Ideal classifier offline experiment}
%\end{figure}

\subsection{Online results}
In order to test our approach in terms of web search engine metrics measuring user satisfaction, we conducted an A/B test \cite{ABTesting}. Some users of Yandex search engine were always presented with ordinary search results (control bucket), which were never diversified with fresh documents, and some users were presented with diversified search results (treatment bucket). We ran the experiment for 13 days in March 2011 and that involved about 10 million queries in each bucket (issued by real users as we filtered out bots and spammers). We measured user satisfaction using metrics suggested by Radlinski et.\ al.\ \cite{Eval}, as they are known to correlate with search result quality. Final results for the control and the treatment are listed in Table \ref{tab:ab}. \par

\begin{table}[ht]
\defaultaddspace 9pt
\caption{\label{tab:ab} User behavior metrics for the control and the treatment buckets}
\begin{center}
\small
\begin{tabular}{p{2.5cm}p{2.9cm}rr}
\toprule
Metric & Meaning & Contr & Treat \\ 
\midrule \addlinespace
Abandonment Rate & \% of queries with no results clicked & 33.65 & 32.77  \\ \addlinespace 
Time to 1st click & Time to first click on any result (in sec) & 10.95 & 10.76 \\ \addlinespace
1st Position CTR & \% of queries with 1st position clicked & 44.22 & 45.31  \\ \addlinespace
2nd Position CTR & \% of queries with 2nd position clicked & 14.88 & 14.92  \\ \addlinespace
1st Click Position &  Position of first click & 1.91 & 1.87 \\
\bottomrule
\end{tabular}
\end{center}
\end{table}

All metrics in the treatment are significantly different from metrics in the control (Mann-Whitney U test, $\alpha=0.01$) and all these differences indicate the increase of search result quality after diversification. In other words, the decrease of abandonment rate means less cases when users could not find any relevant result, the decrease of the mean first click position indicates that top results became more relevant, the decrease of the mean time to first click also indicates that relevant results received higher ranks and hence could be spotted faster, and the increase of CTRs of the URLs at the first two positions also indicates their increased relevance. \par

\section{Conclusions and Future Work}

In this paper, we present an approach to improve recency ranking, while preserving the overall relevance of the ordinary search result. We developed a multi-grade recency sensitive queries classifier that predicts the degree of the need in recent documents. We further demonstrated how to diversify the ordinary search result with fresh documents by maximizing the search quality measure which takes the query's temporal ambiguity into account. We demonstrated the behavior of our diversification model in different cases using a set of judged queries. We finally confirmed the intuitions behind our approaches by a large-scale online experiment involving millions of queries from real users. \par
While we consider a fixed time window to determine if documents are fresh, it definitely makes more sense to determine time window which takes the essence of the information need expressed in the query into account. We also need to more systematically handle the challenge of score normalization to obtain the probabilities of document relevance generated according to each possible definition of relevance. In this regard, we look forward to exploit the techniques of results merging developed in the area of Distributed Information Retrieval \cite{DIR}. \par
The diversification based approach to recency ranking can be also useful to aggregate documents from a set of relevant verticals (videos, images or shopping items). However, the danger of over-diversification is not well studied yet. It is not clear if users would prefer too many results of different kinds blended into one search result page. Our long term goal is to develop a unified approach to deal with several kinds of query ambiguities: topical and non-topical. 

%\section{Acknowledgments}
%Acknowledgments are omitted for submission, so not to compromise anonymity.

%\bibliographystyle{abbrv}
%\bibliography{expert}

\end{document}